\documentclass{ieeeaccess}

\usepackage{amsmath,amsfonts}
\usepackage{algorithmic}
\usepackage{array}
\usepackage{subfigure}
\usepackage[caption=false,font=normalsize,labelfont=sf,textfont=sf]{subfig}
\usepackage{textcomp}
\usepackage{bm}
\usepackage{cite}
\usepackage{amsthm}
\usepackage{amsmath}
\usepackage{amssymb}

\usepackage{stfloats}
\usepackage{url}
\usepackage{verbatim}

\usepackage{graphicx}

\def\BibTeX{{\rm B\kern-.05em{\sc i\kern-.025em b}\kern-.08em
    T\kern-.1667em\lower.7ex\hbox{E}\kern-.125emX}}
\usepackage{balance}
\usepackage{soul}

\usepackage{booktabs} 

\usepackage{cite}
\usepackage{amsmath,amssymb,amsfonts}
\usepackage{algorithmic}
\usepackage{graphicx}
\usepackage{textcomp}
\usepackage{mathtools}
\usepackage{bm}
\usepackage{url}
\usepackage{multirow}
\graphicspath{{./figs/}}

\DeclarePairedDelimiter\abs{\lvert}{\rvert}%

\renewcommand{\Re}{\operatorname{Re}}

\def\BibTeX{{\rm B\kern-.05em{\sc i\kern-.025em b}\kern-.08em
    T\kern-.1667em\lower.7ex\hbox{E}\kern-.125emX}}
\begin{document}
\history{Date of publication xxxx 00, 0000, date of current version xxxx 00, 0000.}
\doi{10.1109/ACCESS.2017.DOI}

\title{A Survey on Noise-Based Communication}
\author{\uppercase{Higo T. P. da Silva}\authorrefmark{1}, 
\uppercase{Hugerles S. Silva\authorrefmark{2} \IEEEmembership{Senior Member, IEEE}, Felipe A. P. Figueiredo\authorrefmark{3}, André A. dos Anjos\authorrefmark{4} and Rausley A. A. Souza\authorrefmark{5}}
\IEEEmembership{Senior Member, IEEE}}
\address[1]{Department of Electrical Engineering, University of Brasilia (UnB), Brasilia, 70.910-900, Brazil (e-mail: higo.silva@unb.br)}
\address[2]{Department of Electrical Engineering, University of Brasilia (UnB), Brasilia, 70.910-900, Brazil (e-mail: hugerles.silva@unb.br)}
\address[3]{National Institute of Telecommunications, Inatel,	Santa Rita do Sapucaí, Brazil (e-mail: felipe.figueiredo@inatel.br)}
\address[4]{Department of Telecommunications,
Federal University of Uberlândia, Patos de Minas - MG, Brazil (e-mail:
andre.anjos@ufu.br)}
\address[5]{National Institute of Telecommunications, Inatel,	Santa Rita do Sapucaí, Brazil (e-mail: rausley@inatel.br)}

\tfootnote{}

\markboth
{Author \headeretal: Preparation of Papers for IEEE TRANSACTIONS and JOURNALS}
{Author \headeretal: Preparation of Papers for IEEE TRANSACTIONS and JOURNALS}

\corresp{Corresponding author: Hugerles S. Silva (e-mail: hugerles.silva@unb.br).}

\begin{abstract}
The proliferation of sixth-generation (6G) networks and the massive Internet of Things (IoT) demand wireless communication technologies that are ultra-low-power, secure, and covert. 
Noise-based communication has emerged as a transformative paradigm that meets these demands by encoding information directly into the statistical properties of noise, rather than using traditional deterministic carriers.
This survey provides a comprehensive synthesis of this field, systematically exploring its fundamental principles and key methodologies, including thermal noise modulation (TherMod), noise modulation (NoiseMod) and its variants, and the Kirchhoff-law-Johnson-noise~(KLJN) secure key exchange. 
We address critical practical challenges such as channel estimation and hardware implementation, and highlight emerging applications in simultaneous wireless information and power transfer~(SWIPT) and non-orthogonal multiple access~(NOMA).
Our analysis confirms that noise-based systems offer unparalleled advantages in energy efficiency and covertness, and we conclude by outlining future research directions to realize their potential for enabling the next generation of autonomous and secure wireless networks.
\end{abstract}


\titlepgskip=-15pt

\maketitle

\section{Introduction} \label{sec:introduction}

The relentless global pursuit of digital transformation is ushering in an era of unprecedented connectivity, marked by the proliferation of massive-scale Internet of Things (IoT) deployments and the ambitious vision for sixth-generation~(6G) wireless networks~\cite{Wang}. 
These future ecosystems envision a seamlessly interconnected world where billions of devices, ranging from miniature sensors and wearable gadgets to autonomous machines, operate continuously in diverse environments, including remote, inaccessible, or security-sensitive locations. 
However, this paradigm shift introduces significant challenges that current wireless communication frameworks find difficult to address.
Paramount among these are the critical demands for ultra-low-power operation to enable years of battery-less or energy-autonomous functionality, stringent requirements for physical-layer security and covertness to protect data and privacy, and the need for resilient operation in spectrally crowded and potentially hostile environments~\cite{Basar_2023}.

Traditional communication schemes, which underpin most contemporary wireless systems, are fundamentally ill-suited for this new frontier. 
Relying on the generation of deterministic, high-power carrier signals for amplitude, frequency, or phase modulation, these methods incur significant energy costs. 
The requisite components, such as oscillators, power amplifiers, and complex mixers, are substantial power drains, making such systems impractical for the stringent energy constraints of pervasive IoT nodes and smart dust applications~\cite{Basar_2023}.
In response, research has increasingly turned towards energy harvesting and simultaneous wireless information and power transfer (SWIPT) as promising avenues to power these devices, allowing them to decode information and scavenge energy from ambient electromagnetic signals~\cite{Hossain}.

Beyond the sheer reduction of power consumption, there is a growing and critical need for covert and resilient communication methods~\cite{Chen}.
Applications spanning secure data transmission, privacy-preserving smart infrastructures, defense, and surveillance necessitate techniques that can operate undetected by adversaries.
Such covertness is achieved when the communication signal is statistically indistinguishable from the ambient background noise, evading conventional detection and interception methods~\cite{Chen}. 
The convergence of these three imperatives (i.e., ultra-low-power operation, energy autonomy, and stealthiness) necessitates a radical departure from traditional design principles, paving the way for a new communication paradigm where noise itself becomes the primary vehicle for information transfer.

The intellectual foundations of this noise-based communication paradigm can be traced back to pioneering work in the early 2000s. L\'{a}szl\'{o} B. Kish introduced the revolutionary concept of zero-signal-power communication, demonstrating that information could be conveyed by modulating the thermal noise generated by resistors at different temperatures or with different resistance values~\cite{Kish}. 
This method, in stark contrast to traditional systems, did not rely on the deliberate transmission of a structured signal, thereby achieving near-zero signal power and offering a fundamental form of physical-layer security. 
Building upon this seminal idea, the Kirchhoff-law-Johnson-noise (KLJN) secure key exchange protocol was developed, leveraging the intrinsic properties of thermal noise and Kirchhoff's circuit laws to enable two parties to establish a shared secret key with information-theoretic security guarantees~\cite{Kish2}.

These early concepts have recently been generalized and formalized into more comprehensive frameworks, significantly expanding their scope and applicability. 
Notable among these are thermal noise modulation (TherMod), which systematically exploits resistor-based thermal noise for data transmission~\cite{Basar_2023}, and the broader concept of noise modulation (NoiseMod), which utilizes artificially generated Gaussian noise waveforms~\cite{Basar_2024}. 
These approaches fundamentally redefine the role of noise in communication systems. 
Instead of treating noise as a detrimental impairment to be mitigated, they encode digital data directly into its statistical properties, specifically, its mean or variance. 
This novel perspective unlocks new forms of communication that are inherently low-power, covert, and compatible with energy harvesting.

The advantages of noise-based communication are manifold and well-documented in the burgeoning literature, including ultra-low transmission power, significant hardware simplification, inherent robustness to jamming and detection, and seamless compatibility with energy harvesting paradigms~\cite{Basar_2023,Basar_2024,Alshawaqfeh,Shen,Yapici2025,Shi,NOMA}. 
Consequently, noise-driven communication schemes are rapidly gaining traction as a viable and disruptive solution to the core challenges posed by next-generation wireless networks~\cite{Basar_2023,Basar_2024}.

Despite the promising progress, a comprehensive and systematic survey that consolidates the fundamental principles, synthesizes the diverse methodologies, critically analyzes performance trade-offs, and maps out the emerging applications and future research directions of this rapidly evolving field is notably absent from the literature. 
This article aims to fill this critical gap. 
In this paper, we provide a comprehensive examination of noise-based communication techniques, highlighting their potential to transform the future landscape of wireless communications. 
We systematically delve into the core principles of various modulation schemes, including TherMod, NoiseMod, and their non-coherent and time-diversity variants, as well as on-off digital noise modulation~(OODN). 
Furthermore, we examine the application of these principles to secure key exchange via the KLJN protocol and its enhancements, address the critical challenge of channel estimation in stochastic signaling environments, discuss practical implementation considerations, and finally, explore integration with emerging technologies like non-orthogonal multiple access (NOMA) and reconfigurable intelligent surface (RIS).
The paper concludes by outlining the key challenges and promising future research directions, underscoring the transformative potential of noise-based systems in enabling ubiquitous, secure, and energy-autonomous wireless networks.

The remainder of this paper is organized as follows.
Section~\ref{noise-based} provides a comprehensive overview of noise-based modulation systems, including TherMod, its variants, and externally generated schemes such as NoiseMod and OODN.
Section~\ref{kljn} discusses secure key exchange mechanisms based on the KLJN protocol and its recent extensions.
Section~\ref{channelestimation} presents channel estimation techniques tailored for thermal noise-based communication.
Section~\ref{practical} addresses implementation and practical considerations, while Section \ref{Sec:Application} explores emerging applications and integration prospects.
Section~\ref{challenges} outlines key challenges, future research directions, and standardization efforts.
Finally, Section~\ref{conclusions} concludes the survey.

\section{Noise-Based Modulation Systems}
\label{noise-based}

Conventional communication systems typically regard noise as a detrimental impairment to be minimized or mitigated. 
However, recent advances have shifted this paradigm by leveraging noise itself as the carrier of information~\cite{Basar_2023,Basar_2024}. 
The fundamental insight behind noise-based communications is that the statistical properties of noise, specifically, its mean and variance, can be modulated to embed and transmit digital data~\cite{Basar_2023}. 

Two principal methods have been proposed for encoding information into noise. 
Concerning the mean-based schemes, the average value of the noise waveform is shifted according to the transmitted information bits. 
A zero-mean Gaussian noise signal may represent the bit \texttt{0}, while a nonzero mean corresponds to the bit \texttt{1}, or vice versa. 
For the variance-based schemes, different symbols are represented by altering the variance of the Gaussian noise process.
For instance, a lower variance may encode a bit \texttt{0} and a higher variance a bit \texttt{1}. 
The variance in these schemes can be estimated via non-coherent energy detection, making them especially suitable for low-complexity receivers.

Noise-based modulation, which incorporates information in the variance, offers advantages over that which encodes information in the mean. 
Firstly, it enables non-coherent detection without requiring precise channel knowledge, which simplifies receiver design and reduces energy consumption~\cite{Basar_2024}. 
Secondly, the use of noise variance as an information-bearing parameter inherently supports covert communication \cite{Basar_2024,Chen}. 
Overall, treating noise as an information-carrying entity, whether by modulating its mean or variance, opens new frontiers for ultra-low-power, resilient, and secure communications, particularly in the context of IoT and 6G networks. 

This section provides an overview of the main noise modulation approaches, describing the system model, detection and binary decision techniques, and performance evaluation. 
The systems considered include the passive TherMod approach and several variants of the active NoiseMod scheme. 
In the channel models discussed below, flat fading is assumed, with independent realizations across symbol intervals. 

\subsection{Thermal Noise Modulation (TherMod)}

The TherMod system is illustrated in Fig.~\ref{fig:thermod_sys}. 
This system exploits the statistical properties of thermal noise to encode digital information, i.e., bits are conveyed by modulating the variance of a transmitted thermal noise signal, achieved via resistor indexing~\cite{Basar_2023}. 
The mean square value of the thermal noise voltage $v_{R}(t)$ across the terminals of a resistor of resistance $R$, denoted by $\mathbb{E}[v_{R}^2(t)]$, is given by~\cite{Basar_2023}
    \begin{equation}
        \label{eq:v_rms}
        \mathbb{E}[v_{R}^2(t)] = 4kTR\Delta f \text{volts}^2,
    \end{equation}
in which $k = 1.38\times 10^{-23}$ J/K is the Boltzmann constant, $T$ is the temperature expressed in Kelvins, and $\Delta f$ is the bandwidth. 
In the TherMod system in particular, two resistors with distinct resistance values, $R_0$ and $R_1$, are associated with bits \texttt{0} and \texttt{1}, respectively. 
Due to the relationship expressed in~\eqref{eq:v_rms}, selecting a resistor effectively alters the variance of the emitted noise waveform~\cite{Basar_2023}. 
The baseband noise waveform is then up-converted, considering a carrier frequency $f_\text{c}$, and transmitted over the wireless channel. 
%
\Figure[t][width=0.98\columnwidth]{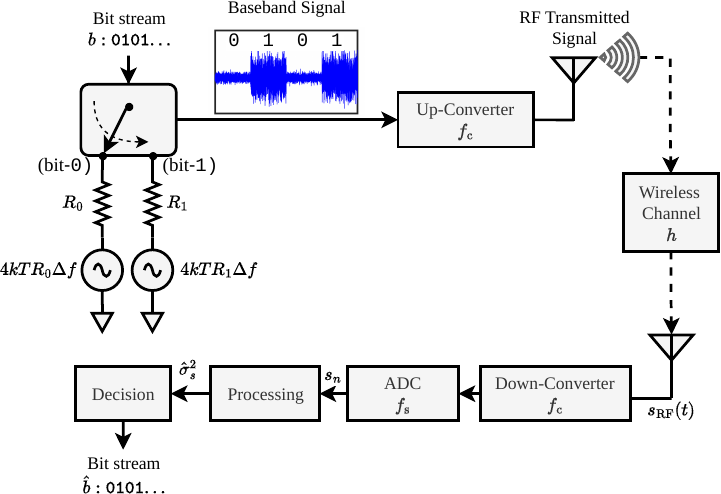}{{TherMod system model illustration.}\label{fig:thermod_sys}}

At the receiver, the passband received radio-frequency (RF) signal $s_{\text{RF}}(t)$ is down-converted to a baseband signal, which is digitized by an analog-to-digital converter (ADC) with sampling frequency $f_\text{s}$. 
The resulting samples are modeled as 
    \begin{equation}
    \label{eq:rec_signal}
     s_n = hr_n + w_n,
    \end{equation}
where $s_n$ denotes the $n$-th complex baseband sample acquired at the receiver, $r_n$ is the zero-mean signal component carrying the intended information via its variance, $h$ represents the small-scale fading coefficient, and $w_n$ is an additive noise component modeled as complex Gaussian noise with zero mean and variance $\sigma_w^2$~\cite{Basar_2023}.
The pure additive white Gaussian noise~(AWGN) channel, without fading effects, can be evaluated simply by equating $h=1$ in~\eqref{eq:rec_signal}. 
According to this model, the variance of $r_n$ depends on the transmitted bit $b \in \{\texttt{0}, \texttt{1}\}$, being $\sigma_0^2$ for $b = 0$ and $\sigma_1^2 = \alpha \sigma_0^2$ for $b = 1$, where $\alpha > 1$ defines the ratio between the high and low noise levels. 
Hence, each bit is associated with either a low- or high-variance noise waveform.

Assuming an extraction of $N$ samples of the received signal per symbol time $T_{\text{b}}$ and quantifying a sampling frequency $f_\text{s}=N/T_{\text{b}}$, the receiver calculates an estimate of the variance by means of~\cite[Eq. (24)]{Basar_2023}
    \begin{equation}
        \label{eq:est_var}
        \hat{\sigma}_s^2 = \frac{1}{N} \sum_{n=1}^{N} |s_n|^2.
    \end{equation}
This estimated variance is then compared to a threshold $\gamma$, and the received bit decision is based on
    \begin{equation}
        \hat{b} = \begin{cases}
            \texttt{0}, & \text{if } \hat{\sigma}_{s}^2 < \gamma, \\
            \texttt{1}, & \text{if } \hat{\sigma}_{s}^2 > \gamma.
        \end{cases}
    \end{equation}
To quantify performance, the ratio between the useful and additive noise variances is applied, which is denoted as $\delta\triangleq\sigma_{0}^2/\sigma_{w}^2$.
Then, the variances of the received samples under each hypothesis become $\tilde{\sigma}_0^2 = \sigma_w^2(1 + \delta)$ and $\tilde{\sigma}_1^2 = \sigma_w^2(1 + \alpha\delta)$. 
Note that $\delta$ is a metric analogous to the signal-to-noise ratio (SNR) used in traditional communication systems. 
Although $\delta = \sigma_{0}^2/\sigma_{w}^2$ represents a ratio between noise powers, the power of the numerator is relative to the noise carrying information, while the power of the denominator is the power of the AWGN term at the receiver.



At the receiver, the detection of transmitted bits relies on the statistical estimation of the variance of the received noise samples. 
Two detection strategies are considered in the literature, namely the energy and the maximum likelihood (ML) detectors.
The first of these is the straightforward non-coherent technique, where the sample variance of the received signal within each bit duration is computed and compared against a decision threshold to infer the transmitted bit. 
In this scheme, the decision threshold must be determined based on the parameters $\alpha$ and $\delta$ and the fading coefficient $h$, which implies that the receiver must have some level of knowledge of the channel~\cite{Basar_2023}.
In turn, the ML detector is an optimal detection approach based on likelihood ratio testing. 
Under the assumption of independent and identically distributed Gaussian noise samples, the ML detection rule compares the estimated variance $\hat{\sigma}^2_s$ against an optimal threshold $\gamma_{\text{opt}}$~\cite{Alshawaqfeh}.
The optimal threshold $\gamma_{\text{opt}}$ is derived analytically based on the variances associated with the two resistors, offering significant performance gains compared to sub-optimal thresholds used in earlier works~\cite{Alshawaqfeh}.


The performance of the TherMod system has been rigorously analyzed under AWGN and Rayleigh fading channels~\cite{Basar_2023}.
Assuming the central limit theorem applied to the sample variance estimation for sufficiently large $N$, the conditioned bit error probability (BEP) to the fading quadratic envelope $\abs{h}^2$ can be written as~\cite[Eq. (29)]{Basar_2023}
    \begin{equation}\label{eq:BEP}
        P_\text{b}\left(\abs{h}^2\right) = Q\left(
            \frac{\sqrt{N} \delta(\alpha - 1)\abs{h}^2}{2 + \delta(\alpha + 1)\abs{h}^2}
        \right),
    \end{equation}
where $Q(\cdot)$ denotes the Q-function.
Note that \eqref{eq:BEP} shows that the system's error performance improves exponentially with the number of samples $N$, particularly when both $\delta$ and $\alpha$ are sufficiently large. 

For the optimal threshold derived in~\cite{Alshawaqfeh}, the average BEP for AWGN channels is given by
    \begin{equation}
    \label{eq:BEP_2}
    P_\text{b} = \frac{1}{2} \left[
    1 - F_{\chi^2_{2N}} \left( \frac{2N}{\sigma_{s_0}^2} \gamma_{\text{opt}} \right)
    + F_{\chi^2_{2N}} \left( \frac{2N}{\sigma_{s_1}^2} \gamma_{\text{opt}} \right)
    \right],
    \end{equation}
where $F_{\chi^2_{2N}}(z)$ represents the cumulative distribution function (CDF) of the chi-squared distribution with $2N$ degrees of freedom, which is expressed as $F_{\chi^2_{2N}}(z) = \gamma_{\text{inc}}\left( N, \frac{z}{2} \right)/\Gamma(N)$, where $\gamma_{\text{inc}}(\cdot, \cdot)$ denotes the lower incomplete gamma function with $\Gamma(\cdot)$ being the gamma function.
Monte-Carlo simulations validate the derived BEP expressions in~\eqref{eq:BEP} and~\eqref{eq:BEP_2}, demonstrating strong agreement between the analytical and simulated curves.




\subsection{P-TherMod}

An extension of the classic TherMod scheme is presented in~\cite{Salem2025}. 
The so-called P-TherMod architecture employs a four-resistor configuration, denoted as $R_{00}$, $R_{01}$, $R_{10}$, and $R_{11}$, which correspond to the noise variances $\sigma_{00}^2$, $\sigma_{01}^2$, $\sigma_{10}^2$, and $\sigma_{11}^2$, respectively. 
This configuration introduces greater diversity in noise variance, providing four distinct levels 
that represent the two-bit sequences \texttt{00}, \texttt{01}, \texttt{10}, and \texttt{11}. 
The variance relationships are defined as
    \begin{equation}
        \sigma_{01}^2 = \alpha \sigma_{00}^2,
    \end{equation}
    \begin{equation}
        \sigma_{10}^2 = \alpha \sigma_{01}^2 = \alpha^2\sigma_{00}^2,
    \end{equation}
and
    \begin{equation}
        \sigma_{11}^2 = \alpha \sigma_{10}^2 = \alpha^3\sigma_{00}^2.
    \end{equation}
Since $\alpha > 1$, it follows that $\sigma_{00}^2 < \sigma_{01}^2 < \sigma_{10}^2 < \sigma_{11}^2$. 
In the P-TherMod scheme, the SNR metric is defined as  $\delta = \sigma_{00}^2 / \sigma_{w}^2$.

The P-TherMod operation follows the same principle as the conventional TherMod scheme. 
After mapping the bit sequences into noisy waveforms, these signals are upconverted to a high-frequency band and transmitted over a wireless channel. 
At the receiver, the noise variance is estimated as in~\eqref{eq:est_var}, and symbol detection is performed using three decision thresholds, $\gamma_{1}$, $\gamma_{2}$, and $\gamma_{3}$, defined as
    \begin{equation}
        \hat{b} = 
        \begin{cases}
            \texttt{00}, & \text{if } \hat{\sigma}_{s}^2 < \gamma_{1}, \\
            \texttt{01}, & \text{if } \gamma_{1} < \hat{\sigma}_{s}^2 < \gamma_{2}, \\
            \texttt{10}, & \text{if } \gamma_{2} < \hat{\sigma}_{s}^2 < \gamma_{3}, \\
            \texttt{11}, & \text{if } \hat{\sigma}_{s}^2 > \gamma_{3}, \\
        \end{cases}
    \end{equation}
where $\gamma_{1} < \gamma_{2} < \gamma_{3}$.
The thresholds are scaled with respect to the AWGN power as $\gamma_{i} = \mu_{i}\sigma_{w}^2$, for all $i \in \{1,2,3\}$, being $\mu_{i}$ constants.

According to~\cite{Salem2025}, the BEP of the P-TherMod scheme, conditioned on the squared magnitude of the channel fading, is given by
    \begin{equation}
        P_\text{b}\left(\abs{h}^2\right) = \frac{1}{12} 
        \sum_{i=1}^{3} \sum_{j=0}^{3}
        \left[
            Q\!\left(
                \frac{\sqrt{N}\,\abs{\omega_{i} - \left(1 - \abs{h}^2 \alpha^{j} \delta\right)}}
                     {1 + \abs{h}^2 \alpha^{j} \delta}
            \right)
        \right],
    \end{equation}
where
    \begin{equation}
        \omega_{i} = 
        \frac{2(1 + \abs{h}^2 \delta)(1 + \abs{h}^2 \alpha^{i} \delta)}
             {2 + \abs{h}^2 \delta \alpha^{i-1} (\alpha + 1)}.
    \end{equation}
Similar to the conventional TherMod, the performance of P-TherMod improves as the number of samples used for variance estimation increases.

\subsection{External Noise Modulation ({NoiseMod})}

NoiseMod refers to a generalized form of noise-based communication where externally generated artificial Gaussian noise is modulated and transmitted~\cite{Basar_2024}. 
Unlike TherMod, which uses thermal noise arising from passive elements such as resistors, NoiseMod uses artificially generated baseband noise samples that are modulated through an in-phase/quadrature (IQ) process to produce a passband noise waveform~\cite{Basar_2024}. 
This is typically implemented using software-defined radio (SDR) platforms or vector signal generators capable of generating and modulating random Gaussian noise. 
Fig.~\ref{fig:noisemod_sys} shows a block diagram of the NoiseMod system.
\Figure[t][width=0.98\columnwidth]{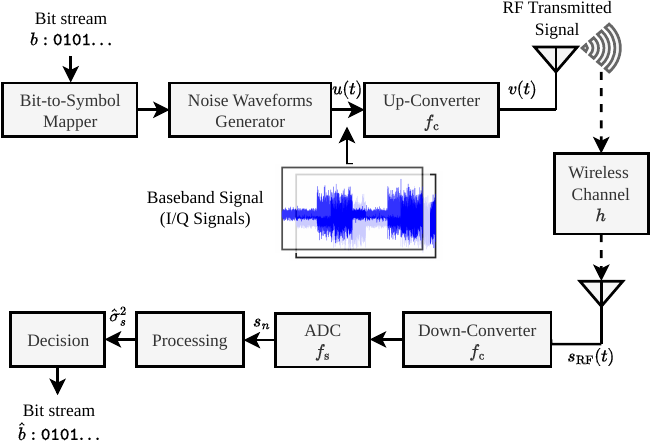}{{NoiseMod system model illustration.}\label{fig:noisemod_sys}}

The transmitted passband waveform $v(t)$ is expressed as $v(t) = \Re\{u(t) e^{j2\pi f_c t}\}$, where $u(t)$ is the complex baseband noise waveform and $\Re\{\cdot\}$ denotes the real part. 
At the receiver, the processing is similar to that of the TherMod system, where the received RF signal $s_{\text{RF}}(t)$ is downconverted to baseband and sampled, yielding discrete-time complex samples for further processing, as expressed in~\eqref{eq:rec_signal}. 
In its basic form, the NoiseMod operation is equivalent to that of the TherMod approach; however, the use of generators provides greater flexibility in controlling the noise variances at the cost of higher power consumption.  
Due to their structural similarity, the conditional BEP of the basic NoiseMod follows the expression in~\eqref{eq:BEP}. 
The average BEP for a given fading scenario is derived by averaging the conditional BEP over the fading distribution.
In~\cite{Basar_2024}, the performance of NoiseMod is analyzed under Rayleigh fading conditions.

\subsection{Non-Coherent NoiseMod}

The non-coherent NoiseMod (NC-NoiseMod) variant addresses a critical limitation of NoiseMod, namely its reliance on channel knowledge for optimal detection. 
In NC-NoiseMod, a mid-bit variance flip is introduced to enable non-coherent detection without requiring channel state information~(CSI)~\cite{Basar_2024}. 
Specifically, the variance of the transmitted noise is switched halfway through each bit duration.
On the one hand, for the bit \texttt{0}, a low-variance noise is transmitted in the first half and a high-variance noise in the second half. 
On the other hand, for the bit \texttt{1}, the order is reversed. 
Fig.~\ref{fig:nc_noisemod} shows a comparison of the waveforms of NoiseMod and NC-NoiseMod.
\Figure[t][width=0.98\columnwidth]{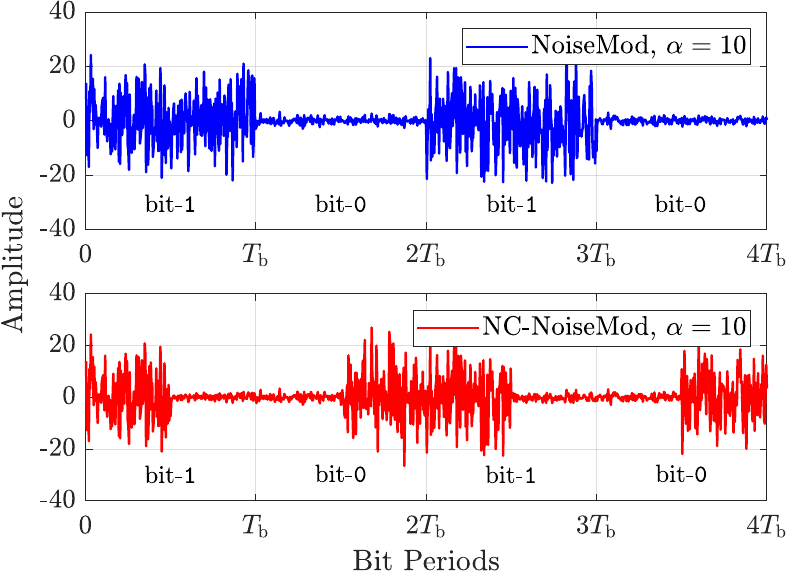}{{Comparison between NoiseMod and NC-NoiseMod waveforms. \label{fig:nc_noisemod}}}

At the receiver, two separate sample variances $\hat{\sigma}^2_{s,1}$ and $\hat{\sigma}^2_{s,2}$ are estimated for the first and second halves of the bit duration as
    \begin{equation}
    \hat{\sigma}^2_{s,1} = \frac{2}{N} \sum_{n=1}^{N/2} |s_n|^2
    \end{equation}
and
    \begin{equation}
    \hat{\sigma}^2_{s,2} = \frac{2}{N} \sum_{n=N/2+1}^{N} |s_n|^2,
    \end{equation}
respectively.
The decision rule is given by
    \begin{equation}
        \label{eq:bit_dec}
        \hat{b} = \begin{cases}
            \texttt{0}, & \text{if } \hat{\sigma}^2_{s,1} < \hat{\sigma}^2_{s,2}, \\
            \texttt{1}, & \text{if } \hat{\sigma}^2_{s,1} \geq \hat{\sigma}^2_{s,2}.
        \end{cases}
    \end{equation}
Therefore, the NC-NoiseMod decision does not require knowledge of the $\alpha$ and $\delta$ parameters nor a definition of a fading-dependent threshold $\gamma$, eliminating the need for CSI and providing significant practical advantages in dynamic or unknown environments.

In~\cite{Basar_2024}, it is demonstrated that the NC-NoiseMod BEP conditioned on the quadratic envelope of the channel fading is expressed by
    \begin{equation}
        P_\text{b}\left(\abs{h}^2\right) = Q\left( \frac{\sqrt{N/2} \abs{h}^2 \delta (\alpha - 1) }{(1 + \abs{h}^2 \delta\alpha)^2 + (1 + \abs{h}^2 \delta)^2} \right).
    \end{equation}
Performance evaluation under Rayleigh fading conditions shows that NoiseMod and NC-NoiseMod achieve comparable numerical performance~\cite{Basar_2024}. 
However, NC-NoiseMod is more advantageous, as it does not require CSI for symbol detection.

\subsection{Time-Diversity NoiseMod}

Time-diversity NoiseMod (TD-NoiseMod) is proposed to enhance the robustness of NoiseMod against small-scale fading effects~\cite{Basar_2024}. 
In TD-NoiseMod, each bit period of $N$ samples is divided into $M$ slots, where $M$ determines the temporal spreading ratio. 
The idea is to spread the samples of the same bit across different fading realizations to exploit time diversity.

Formally, if each bit duration is assigned $N$ samples, then these are divided into $M$ equal groups, each transmitted at different times. 
The receiver then estimates the overall variance by aggregating the variances from these different slots, as
    \begin{equation}
        \sigma_s^2 =
        \begin{cases}
            \displaystyle \frac{1}{M}\sum_{m=1}^M \sigma_w^2 (1+|h_m|^2\delta), & \text{for bit 0}, \\
            \displaystyle \frac{1}{M}\sum_{m=1}^M \sigma_w^2 (1+|h_m|^2\alpha\delta), & \text{for bit 1},
        \end{cases}
    \end{equation}
where $h_m$ represents the independent fading coefficient for the $m$-th slot. 
The detection phase of TD-NoiseMod follows a similar fashion to NoiseMod, with the binary decision determined by comparing the estimated variance with a predefined threshold.

According to~\cite{Basar_2024}, the TD-NoiseMod conditioned BEP is expressed by
    \begin{equation}\label{eq:TD-BEP}
        P_\text{b}\left(\abs{g}^2\right) = Q\left(
            \frac{\sqrt{N} \delta(\alpha - 1)\abs{g}^2}{2 + \delta(\alpha + 1)\abs{g}^2}
        \right),
    \end{equation}
where $\abs{g}^2 = \sum_{m=1}^{M}\abs{h_{m}}^2$. 
It can be seen that when $M=1$, the BEP of the TD-NoiseMod in~\eqref{eq:TD-BEP} is reduced to the BEP of the basic NoiseMod. 
As shown in~\cite{Basar_2024}, under Rayleigh fading, the unconditional BEP of TD-NoiseMod can be determined by numerically integrating over the chi-square distribution of $\abs{g}^2$ with $2M$ degrees of freedom.
Results indicate that TD-NoiseMod attains a diversity order proportional to $M$, with its numerical BEP approximately following the reference behavior of $10^{1-M}/(N\delta^{M})$ across the evaluated $N$ and $M$ values.

\subsection{On-Off Digital Noise Modulation}

OODN modulation is a variant of NoiseMod inspired by the simplicity of on-off keying (OOK) modulation~\cite{Anjos}. 
In this scheme, bit \texttt{0} is represented by a completely clipped transmission, where $r_{n}=0$ for all samples in the bit period. 
In turn, bit \texttt{1} is represented by a zero-mean complex noise with variance $\sigma_{1}^2$. 
Consequently, when bit \texttt{0} is transmitted, the received samples correspond only to the receiver's AWGN component, while at bit $\texttt{1}$, the received signal is $s_{n} = hr_{n} + w_{n}$, with $r_{n}$ having variance $\sigma_{1}^2$. 

Using the $N$ received samples, the receiver estimates the per-bit variance $\hat{\sigma}_{s}^2$ as defined in~\eqref{eq:est_var}, and compares this estimate with a threshold $\gamma$ to determine the transmitted bit following~\eqref{eq:bit_dec}. 
In~\cite{Anjos}, the authors define the threshold $\gamma$ based on the statistical properties of $\sigma_{1}^2$ and $\sigma_{w}^2$. 
It is demonstrated that the conditioned BEP of the OODN modulation can be expressed by
    \begin{equation}
        \label{eq:oodn_bep}
        P_\text{b}\left(\abs{h}^2\right) = Q\left(\sqrt{N} \frac{2\abs{h}^2\bar{\delta}}{1 + 2\abs{h}^2\bar{\delta}}\right),
    \end{equation}
in which $\bar{\delta} = \sigma_{1}^2/\sigma_{w}^2$. The performance of OODN is evaluated in~\cite{Anjos} over both AWGN and fading channels modeled by the $\kappa$-$\mu$ distribution, a model that captures line-of-sight (LoS) and non-line-of-sight (NLoS) environments. 
The results clearly demonstrate that, for all evaluated cases, OODN yields superior performance compared to NoiseMod under the same SNR conditions and for any admissible values of $\alpha$ and $\delta$.

\subsection{Noise-Domain Non-Orthogonal Multiple Access}

In traditional orthogonal multiple access schemes, different users are 
mutually separated by allocating orthogonal resources such as time or frequency. 
In contrast, non-orthogonal multiple access (NOMA) enables multiple users to 
share the same resources simultaneously. 
In~\cite{NOMA}, a scheme called noise-domain NOMA (ND-NOMA) is proposed, which extends noise-based 
communication concepts to a NOMA framework. 
In the system model, a base station (BS) establishes uplink and downlink communications with two users, $\text{U}_{1}$ and $\text{U}_{2}$, while both share the same time-frequency resources. 

Let the mean and variance of the transmitted signals from $\text{U}_{1}$ and $\text{U}_{2}$ be denoted as $(m_{1}, \sigma_{1}^2)$ and $(m_{2}, \sigma_{2}^2)$, respectively. 
In the uplink phase, $\text{U}_{1}$ encodes its bits by modulating the mean of a noisy signal, such that $m_{1} \in \{m_{1,l}, m_{1,h}\}$, while keeping the variance fixed. 
The signals with means $m_{1,l}$ and $m_{1,h}$ represent bits \texttt{0} and \texttt{1}, respectively. 
Conversely, $\text{U}_{2}$ encodes its bits by maintaining a zero-mean signal and varying the noise variance, 
$\sigma_{2}^2 \in \{\sigma_{2,l}^2, \sigma_{2,h}^2\}$, where the lower and higher variances correspond to bits \texttt{0} and \texttt{1}, respectively. 
The BS receives a superposition of both users’ signals that is subsequently sampled.
The $n$-th sample of the received signal in the uplink is written as
    \begin{equation}
        s_{n} = h_{1}r_{1;n} + h_{2}r_{2;n} + w_{n},
    \end{equation}
in which $h_{1}$ and $h_{2}$ are the complex channel coefficients for $\text{U}_{1}$ and $\text{U}_{2}$ links, and $r_{1;n}$ and $r_{2;n}$ denote the samples of the transmitted signals of $\text{U}_{1}$ and $\text{U}_{2}$, respectively.
Based on these samples, the BS performs a sequential detection: $\text{U}_{1}$’s bit is decoded using a minimum-distance detector based on mean estimation, while $\text{U}_{2}$’s bit is decoded using a threshold-based detector that relies on variance estimation. 

Similarly, in the downlink phase, the BS transmits a Gaussian signal, which is denoted as $r_{\text{BS}}$, whose mean and variance respectively encode the bits intended to $\text{U}_{1}$ and $\text{U}_{2}$, respectively. 
The samples of the received signal at $\text{U}_{1}$ and $\text{U}_{2}$ are respectively expressed by
    \begin{equation}
        s_{n;1} = h_{1}r_{\text{BS};n} + w_{1;n},
    \end{equation}
and
    \begin{equation}
        s_{n;2} = h_{2}r_{\text{BS};n} + w_{2;n},
    \end{equation}
in which $r_{\text{BS};n}$ is the $n$-th sample of $r_{\text{BS}}$ and $w_{1;n}$ and $w_{2;n}$ are AWGN samples at the corresponding receivers.
Upon reception, $\text{U}_{1}$ decodes its bit using a minimum-distance detector based on mean estimation, whereas $\text{U}_{2}$ decodes its bit using a threshold-based detector that relies on variance estimation. 
Fig.~\ref{fig:nd_noma} illustrates the ND-NOMA system in both uplink and downlink operations.
\Figure[t][width=0.95\columnwidth]{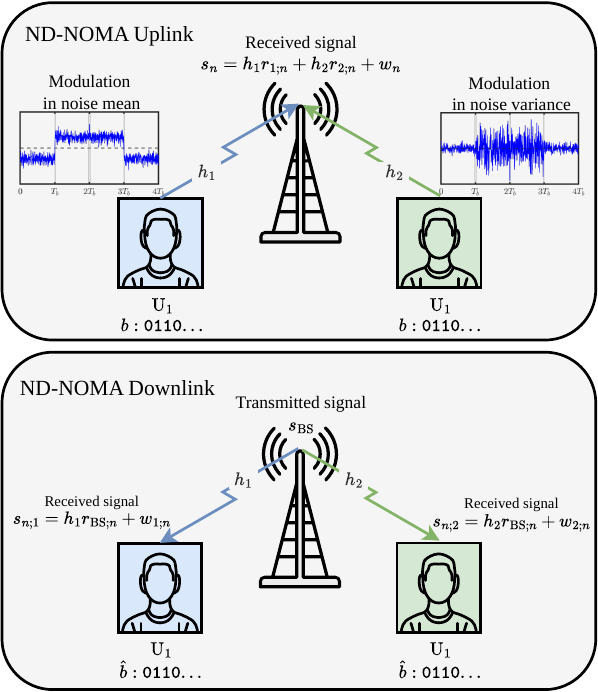}{{Uplink and downlink ND-NOMA illustration.}\label{fig:nd_noma}}

The authors in~\cite{NOMA} evaluate the performance of ND-NOMA in terms of the BEP for each user, considering both uplink and downlink scenarios. 
In their analysis, the channel envelope is modeled as Rician, enabling performance evaluation under both LoS and NLoS conditions by adjusting the Rician factor. 

\subsection{Performance Remarks}

In terms of comparative performance, the TherMod system exhibits the poorest BEP and offers the least flexibility in controlling the variances of the noisy signal. 
Nevertheless, it is highly energy-efficient due to its passive nature. 
Additionally, its performance can be significantly enhanced by using an optimal detection threshold. 
The P-TherMod configuration outperforms the conventional TherMod scheme, achieving a lower BEP under equivalent conditions. 
Moreover, P-TherMod enables a higher data rate by transmitting two-bit sequences per signaling interval. 
These performance gains, however, come at the cost of increased detection complexity, resulting from the greater diversity of noise variances.

The NoiseMod and NC-NoiseMod schemes achieve numerically equivalent bit error performance. 
However, NC-NoiseMod demonstrates greater robustness in practical scenarios, as it does not require CSI for detection. 
As a disadvantage, NC-NoiseMod does not provide diversity. 
In contrast, the TD-NoiseMod scheme offers a diversity order proportional to the number of temporal spreading slots, although it requires CSI for proper operation.
On the other hand, according to the results reported in~\cite{Anjos}, the OODN scheme can outperform NoiseMod in both LoS and NLoS environments, while also exhibiting lower power consumption due to signal clipping at zero bits.

In turn, the ND-NOMA is a multiple-access scheme that operates in the noise domain, significantly reducing power consumption and system complexity. 
The results presented in~\cite{NOMA} indicate that ND-NOMA achieves low BEP values, making it highly suitable for next-generation IoT networks. 
Theoretical analyses and computer simulations demonstrate that ND-NOMA attains exceptionally low BER values in both uplink and downlink transmissions, even under Rician fading conditions.
By leveraging the inherent properties of noise, ND-NOMA offers a promising platform for the long-term deployment of low-cost and low-complexity IoT devices.

Overall, each noise-based system presents distinct advantages and disadvantages, revealing clear trade-offs between performance and energy efficiency.
Moreover, there are inherent trade-offs among the number of samples, reliability, and data rate. 
As discussed throughout this section, the BEPs of the noise modulation schemes are inversely proportional to the number of samples per symbol interval. Specifically, increasing $N$ enhances communication reliability; however, for a fixed sampling frequency $f_\text{s}$, a higher $N$ reduces the data rate and increases system latency. 
Consequently, system designers must carefully select $N$ based on application requirements: low-latency systems may favor smaller $N$, whereas ultra-reliable systems may tolerate lower data rates to achieve higher robustness.


\section{KLJN Secure Key Exchange}
\label{kljn}

This section presents the KLJN secure key exchange~ scheme, which enables the secure transfer of binary symbols through the use of redundancy with noisy signals. Building upon the classic formulation introduced by Kish~\cite{Kish2}, several variant approaches for secure key exchange have been proposed~\cite{Tasci}, which are also discussed in this section.

\subsection{Classical KLJN Scheme}

The KLJN uses thermal noise generated by two resistors for secure key exchange between two entities (Alice and Bob)~\cite{Kish2}.
In this design, Alice and Bob, who are connected by a wired channel, have a low-valued resistor $R_L$ and a high-valued resistor $R_H$.
During each transmission interval, both parties randomly select one resistor and connect it to the common wire channel for communication. Depending on the resistor selections, three distinct noise power levels can be observed on the channel, namely: (i) low noise level (bits-\texttt{00}), where both parties select $R_L$; (ii) high noise level (bits-\texttt{11}), in which both parties select $R_H$ and (iii) intermediate noise level (bits-\texttt{01} or bits-\texttt{10}), where one party selects $R_L$ and the other selects $R_H$~\cite{Kish2}. This binary exchange scheme is illustrated in Fig.~\ref{fig:kljn_ske}.
%
\Figure[t][width=0.98\columnwidth]{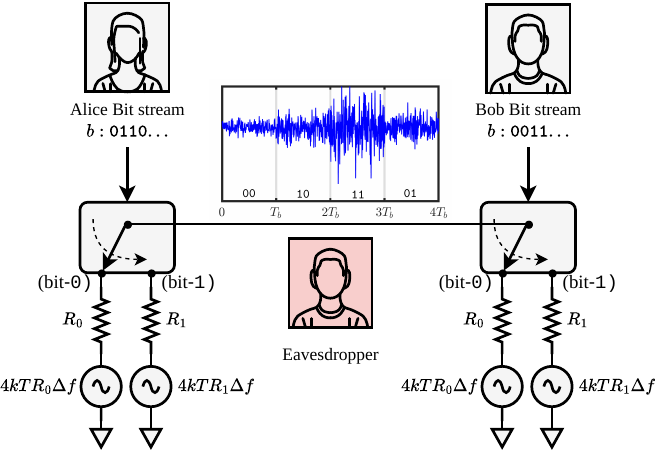}{{KLJN secure key exchange scheme diagram.}\label{fig:kljn_ske}}

In cases \texttt{00} and \texttt{11}, an eavesdropper can determine Alice’s and Bob’s bits from noise variance measurements, due to the one-to-one correspondence of these sequences. This situation represents an insecure binary swap. Therefore, these binary keys are discarded. In contrast, cases \texttt{01} and \texttt{10} generate noise with identical variances, making it impossible for the eavesdropper to distinguish which side transmitted bits \texttt{0} or \texttt{1}. This mechanism can be interpreted as a simple yet effective encryption protocol, providing security against various forms of attack~\cite{Basar_2023}. Meanwhile, Alice and Bob can decode each other’s messages since each party knows their own bit. In this scheme, the bit itself functions as a decryption key, revealing the other participant’s information~\cite{Basar_2023}. In~\cite{Tasci}, it is presented some enhanced variants and detectors for the classical KLJN scheme. 

The reliability performance of the KLJN binary exchange is evaluated in~\cite{Basar_2023}, where a threshold-based detection for the noise variance is applied, considering two threshold values $\gamma_{1}$ and $\gamma_{2}$. Denoting the three variances for the low, intermediary and high states as $\sigma_{00}^2$, $\sigma_{01}^2$ and $\sigma_{11}^2$, the thresholds are defined as $\gamma_{1} = \beta \sigma_{00}^2$ and $\gamma_{2} = \kappa \sigma_{00}^2$, with  $\sigma_{00}^2 < \gamma_{1} < \sigma_{01}^2 < \gamma_{2} < \sigma^{2}_{11}$ and $\beta > 1$  and $\kappa > 1$ being constants. In this context, the BEP for the AWGN channel of the classical KLJN scheme is expressed as~\cite{Basar_2023}
\begin{equation}
\begin{aligned}
    P_\text{b}^{\text{KLJN}} = &\frac{1}{4}\Bigg[ Q\left(\frac{\beta - 1}{\sqrt{2/N}}\right) + Q\left(\frac{\alpha - \kappa}{\sqrt{2/N}}\right) \\
    &+ Q\left(\frac{2\alpha - \beta - \beta\alpha}{2\alpha\sqrt{2/N}}\right) + Q\left(\frac{\kappa + \kappa\alpha - 2\alpha}{2\alpha\sqrt{2/N}}\right)\Bigg],
\end{aligned}
\end{equation}
in which $1 < \beta < \frac{2\alpha}{1 + \alpha} < \kappa < \alpha$.

\subsection{Flip-KLJN}

To enhance the efficiency and utilization of KLJN, the Flip-KLJN protocol is introduced in~\cite{Tasci}. Flip-KLJN employs a dynamic strategy in which the resistor-bit mappings are flipped based on the exchanged random bits, thus enabling the secure use of all noise levels, including the previously discarded low (\texttt{00}) and high (\texttt{11}) levels in the classical KLJN technique~\cite{Tasci}. 

The Flip-KLJN scheme works as follows. During normal operation, the mapping of resistors to logical bits is conventional: $R_L$ maps to bit \texttt{0} and $R_H$ to bit \texttt{1}. Upon encountering a specific binary pattern, such as sequence \texttt{10}, both parties simultaneously flip their resistor-bit mapping: $R_H$ now maps to \texttt{0} and $R_L$ to \texttt{1}. This flipping operation is triggered only when the intermediate noise level is observed, ensuring that the eavesdropper remains unaware of the state transitions. By introducing randomized state transitions without revealing any deterministic pattern, Flip-KLJN effectively confuses the potential eavesdropper and achieves unconditional security across all possible noise levels~\cite{Tasci}. Furthermore, because the \texttt{00} and \texttt{11} cases are now securely utilized, the effective key generation rate is doubled compared to the classical KLJN scheme.

The error rate of the Flip-KLJN is evaluated in~\cite{Tasci}, where it is demonstrated that the closed-form BEP under an AWGN channel is calculated by
\begin{equation}
    P_\text{b}^{\text{Flip-KLJN}} = 3P_\text{b}^{\text{KLJN}} - 2(P_\text{b}^{\text{KLJN}})^2.
\end{equation}

\subsection{P-TherMod KLJN}

Let the resistors at Alice's and Bob's sides be denoted by $R_{A;\text{p}}$ and $R_{B;\text{p}}$, respectively, with $\text{p} \in \{00,01,10,11\}$. It is defined that Alice’s three smallest resistances correspond to Bob’s three largest, that is, $R_{A;\text{00}} = R_{B;\text{01}}$, $R_{A;\text{01}} = R_{B;\text{10}}$, and $R_{A;\text{10}} = R_{B;\text{11}}$. 
Additionally, it is assumed that the resistances of Alice and Bob satisfy $R_{A;\text{p}} = \alpha R_{B;\text{p}}$. By applying the P-TherMod four-resistor configuration, 
a total of 16 distinct 4-bit sequences can be transmitted, 
where the two most significant bits represent Alice’s information 
and the two least significant bits represent Bob’s. 
Note that each binary sequence is mapped to a noise variance $\sigma^2_{\text{pp'}}$, 
with $\text{p},\text{p'} \in \{00,01,10,11\}$ denoting Alice and Bob’s respective bit pairs. 
Each side decodes the other’s message using voltage and current measurements, 
together with its own transmitted sequence. 
Noise variance detection is performed through a threshold-based method employing nine thresholds, 
denoted as $\gamma_{i}$, $i \in \{1,\ldots,9\}$. 
These thresholds are scaled with respect to the lowest variance $\sigma^{2}_{0000}$ as 
$\gamma_{i} = \beta_{i}\sigma^{2}_{0000}$. 

The exact expression for the BEP of the P-TherMod KLJN scheme over AWGN channels is provided in~\cite{Salem2025}. 
Due to its complexity, it is omitted here; however, a simplified form is presented for moderate values of $N$ and $\alpha$ (e.g., $N=100$ and $\alpha=10$), which reduces the BEP in AWGN channels to
\begin{equation}
\begin{aligned}
    P_\text{b} = \frac{1}{48}\Bigg[ 
        &\sum_{i=1}^{3} 
        Q\!\left(
            \sqrt{\frac{N}{2}}\,
            \frac{\alpha^{3}(1 + \alpha) - \beta_{i}(1 + \alpha^{3})}
                 {\alpha^{3}(1 + \alpha)}
        \right) \\
        &+ 
        \sum_{i \in \{1,3,5\}} 
        Q\!\left(
            \sqrt{\frac{N}{2}}\,
            \frac{\alpha^{2} - \beta_{i}}{\alpha^{2}}
        \right)
    \Bigg].
\end{aligned}
\end{equation}

\subsection{Performance Remarks}

In terms of security in binary sequence exchange, the KLJN scheme has demonstrated 
strong resilience against a wide range of attack strategies, including those based on 
cable resistance, temperature imbalance in the communication channel, finite propagation time, 
directional coupler analysis, second-law violations, transient effects, and current injection 
attacks~\cite{Vadai2016, Kish2016, Gunn2015}. Regarding reliability, errors primarily result from inaccuracies in noise variance estimation due to the finite number of samples used in detection. Similarly, the Flip-KLJN variant introduces additional challenges, such as synchronization mismatches during the state-flipping process, which can lead to error propagation. From a security perspective, both schemes provide the same level of secrecy; however, Flip-KLJN achieves higher key generation efficiency since it exploits all possible binary state combinations. In contrast, the P-TherMod-based approach yields a higher throughput and superior BEP performance compared to the conventional KLJN scheme, due to its ability to handle a greater number of binary sequences.

All KLJN-based schemes benefit from an increased number of samples per bit interval, as the BEP decreases with growing $N$. The Flip-KLJN scheme, however, exhibits a higher BEP than the classical KLJN, thus requiring a larger number of samples to achieve comparable error rates. This degradation is primarily attributed to error propagation caused by state switching. Despite its higher BEP, Flip-KLJN attains a superior key rate by exploiting all possible state combinations. Furthermore, the integration of advanced detection techniques, such as the joint voltage-current detector (JVCD), significantly enhances the BER performance of both KLJN and Flip-KLJN systems. The JVCD improves detection robustness by simultaneously processing voltage and current measurements, thereby mitigating decision errors under noisy conditions. Overall, while Flip-KLJN introduces a slight degradation in BER performance relative to the classical KLJN scheme, its ability to double the key rate while maintaining unconditional security offers substantial practical advantages, particularly for secure and energy-efficient key distribution over classical physical channels.

\section{Channel Estimation via Thermal Noise}
\label{channelestimation}

Channel estimation is essential in TherMod systems to achieve reliable communication under fading conditions. 
Unlike conventional systems, TherMod does not transmit deterministic pilot sequences. 
Instead, it uses random thermal noise samples whose statistics are modulated for information transfer. 
Therefore, classical channel estimation techniques need to be adapted. 
In~\cite{Shen}, a channel estimation framework is presented for noise-based communications, where the channel coherence time supports the transmission of a total of $N_\text{b}$ bits, where each bit is represented by $N$ noise samples. 
Therefore, the block supports a total of $N\times N_\text{b}$ samples. 
The frame is divided into two phases: a pilot transmission phase consisting of $N_\text{p}$ known bits and a data transmission phase containing $(N_\text{b} - N_\text{p})$ bits, as illustrated in Fig.~\ref{fig:ch_est}.
\Figure[t][width=0.98\columnwidth]{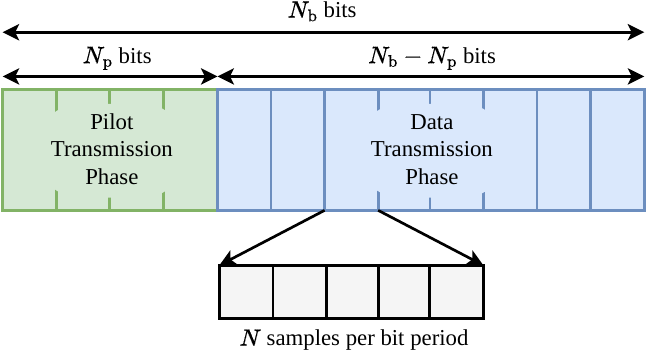}{{Transmission frame for channel estimation.}\label{fig:ch_est}}

During the $j$-th bit period, $j=1,2,\dots,N_\text{b}$, the received signal $\mathbf{y}_{j} \in \mathbb{C}^{N\times 1}$ can be expressed as
    \begin{equation}
        \mathbf{y}_{j} = h\mathbf{r}_{j} + \mathbf{w}_{j},
    \end{equation}
in which $\mathbf{r}_{j} \in \mathbb{C}^{N\times 1}$ and $\mathbf{w}_{j} \in \mathbb{C}^{N\times 1}$ are the transmitted noise sequence and the AWGN vector at the receiver, respectively. 
The elements of $\mathbf{r}_{j}$ and $\mathbf{w}_{j}$ are independently and identically distributed zero-mean circularly symmetric complex Gaussian random variables with variances $\sigma_{\text{i},j}^2$, with $\text{i} \in \{0,1\}$, and $\sigma_{w}^2$, respectively, in which $\sigma_{0,j}^2 = \sigma_{0}^2$ and $\sigma_{1,j}^2 = \sigma_{1}^2$, for any value of $j$. 
In the pilot phase, the estimation is made based on the $N_\text{p}$ received pilots $\mathbf{y}_\text{p} = [\mathbf{y}_{1}^\text{T},\mathbf{y}_{2}^\text{T},\cdots,\mathbf{y}_{N_\text{p}}^\text{T}]^{T}$, with $(\cdot)^\text{T}$ denoting the transposition operation.

The Cramér-Rao bound (CRB) provides a theoretical lower bound on the variance of any unbiased estimator of the channel squared envelope $\Omega = |h|^2$. 
Considering the described pilot phase, the log-likelihood function of the received pilot signals $\textbf{y}_\text{p}$ can be expressed as a product of Gaussian densities.
Deriving the Fisher information leads to a CRB of the squared channel envelope estimation $\hat{\Omega}$ written as~\cite[Eq. (5)]{Shen}
    \begin{equation}    \label{eq:ml_crb}
    \text{CRB}(\hat{\Omega}) = \frac{1}{\displaystyle\sum_{j=1}^{N_\text{p}} \frac{\sigma_{\text{i},j}^2N}{(\Omega \sigma_{\text{i},j} + \sigma_w)^2}}.
    \end{equation}
Note that the CRB decreases as the number of pilot bits $N_\text{P}$ increases and the power of the pilot noise samples $\sigma_{\text{i},j}^2$ increases. Based on~\eqref{eq:ml_crb}, the authors of ~\cite{Shen} propose the use of all-one or all-zero pilots.

\subsection{Modified ML Channel Estimator}

The conventional ML estimator for $\Omega$, derived by maximizing the likelihood function, suffers from feasibility issues when the received pilot power is small.
To overcome this, a modified ML estimator is proposed in~\cite{Shen}. 
The core idea is to augment the ML cost function with a log-barrier penalty, such as~\cite[Eq. (9)]{Shen}
    \begin{equation}
    \min_{\Omega} \, \varphi(\Omega) = f(\Omega) - \xi \ln(\Omega),
    \end{equation}
where $f(\Omega)$ is the original ML objective and $\xi > 0$ is a penalty factor that enforces $\Omega > 0$.

The modified ML closed-form optimal solution is derived as~\cite[Eq. (10)]{Shen}
    \begin{align}
    \Omega^\star &= \frac{\| \textbf{y}_\text{p} \|^2 + 2 \sigma_w \xi - \sigma_w N N_\text{p}}{2 \sigma_{1} (N N_\text{p} - \xi)}\nonumber\\
    &+ \frac{\sqrt{4 \sigma_w \xi \| \textbf{y}_\text{p} \|^2 + \left(\| \textbf{y}_\text{p} \|^2 - \sigma_w N N_\text{p}\right)^2}}{2 \sigma_1 (N N_\text{p} - \xi)}.
    \end{align}

\subsection{Performance Remarks}

The interplay between mean squared error (MSE), CRB, and BER provides valuable insights into the quality of channel estimation and its impact on overall system performance. 
While the CRB values establish a theoretical lower limit for the variance of any unbiased estimator, the MSE ones reflect the practical accuracy of the proposed estimation scheme. 
In turn, BER values directly quantify the effect of estimation quality on performance. 

Extensive simulations presented in~\cite{Shen} confirm the effectiveness of the proposed modified ML estimator under imperfect CSI conditions. 
The MSE of the estimator closely follows the CRB across a wide range of SNR, expressed by the parameter $\delta$, indicating near-optimal performance. 
Furthermore, all-one pilot sequences consistently outperform all-zero sequences in terms of MSE, especially at low SNRs, aligning with the theoretical predictions from CRB analysis. 
When the estimated CSI value is used in a non-coherent energy detection receiver, the resulting BER performance closely matches that obtained with perfect CSI, even when using a small number of pilot bits (e.g., $N_\text{p} = 4$). 
Increasing the number of pilots further enhances both MSE and BER, confirming the robustness of the proposed estimation approach and its suitability for practical TherMod systems targeting ultra-low-power wireless communication.

\section{Implementation and Practical Considerations} \label{practical}

Noise-based communication promises ultra-low-power and covert wireless transmission by embedding information into the statistical properties of noise. 
However, translating theoretical designs into practical systems requires careful consideration of hardware components, receiver architecture, power management, and signal fidelity.

\subsection{TherMod}    \label{subsec:natural}
TherMod transmitters benefit from a highly simplified hardware structure. 
Information bits are modulated by switching between high and low resistance values, which generate thermal noise with distinct variances. 
This process is achieved using low-power electronic switches. 
High-gain directional antennas, such as horn antennas, are recommended to focus the weak noise signals toward the receiver.
In particular, in stealth operation modes, the transmitter does not require RF oscillators, mixers, or amplifiers, significantly reducing complexity, cost, and energy consumption.

On the receiver side, two main detection strategies can be adopted. 
Non-coherent energy detection estimates the noise variance over each bit period and compares it to a threshold, offering simplicity, low power consumption, and excellent performance in many practical cases. 
Alternatively, coherent ML detection, which relies on pilot-assisted channel estimation, offers higher accuracy in fading conditions or when utilizing external noise sources, albeit at the expense of increased complexity and energy consumption.

The power budget and the fidelity of the noise generation are critical factors. 
In passive implementations, Johnson-Nyquist noise from resistors enables near-zero transmit power. 
For an extended communication range, external noise generators or amplifiers may be used, increasing energy demand. 
In such cases, care must be taken to preserve the Gaussian nature of the signal and avoid spectral artifacts. 
On the receiver side, accurate measurement of noise statistics requires sufficient sampling rates, proper bandwidth control to ensure whiteness, and calibration to minimize hardware mismatches. 
Although TherMod systems are conceptually simple, their successful deployment depends on addressing practical design and implementation challenges.

\subsection{NoiseMod, NC/TD variants, and OODN} \label{subsec:artificial}
For artificial noise-based modulations such as NoiseMod, its NC and TD variants, and OODN, the transmitter can be realized on software-defined radios (SDRs) or field-programmable gate array/system-on-chip (FPGA/SoC) platforms that generate complex baseband Gaussian samples with controlled variance per symbol window. 
Practical pipelines pair multiple pseudo-random number generator (PRNG) lanes (e.g., linear feedback shift register, LFSR, or permuted congruential generator, PCG) with lightweight “Gaussianizers” (sum-of-uniforms, lookup-table (LUT) remapping, or the Ziggurat method \cite{Sileshi2014}) and a fixed-point scaler to set each symbol’s variance, or to toggle on/off in OODN.

Spectral compliance can be ensured by shaping and confining the noise with a baseband low-pass or root-raised-cosine filter (LPF/RRC), ramping variance at symbol boundaries to limit spectral splatter, and calibrating IQ gain and phase, as well as DC offsets to maintain a noise-like spectrum. 
Oversampling at the digital-to-analog converter (DAC), together with a moderate effective number of bits (ENOB) of about $12-14$ bits (balancing dynamic range and quantization noise without increasing power or chip area) and an analog anti-alias filter, preserves in-band whiteness with minimal power overhead. 
The sampling frequency \(f_\text{s}\) sets the usable noise bandwidth (\(B\!\approx\! f_\text{s}\) for complex baseband after filtering) and, together with the window length \(N\), determines the symbol/bit duration \(T_\text{b}=N/f_\text{s}\) and thus the bit rate \(R_\text{b}=f_\text{s}/N\); increasing \(f_\text{s}\) expands bandwidth and potential throughput (or processing margin), while increasing \(N\) improves reliability at the expense of rate and latency. 
Frame structures that carry “statistical pilots” (known low/high-variance or silence/active segments) can be used for self-calibration, clock alignment, symbol synchronization, CSI estimation, and drift tracking.

Decisions at the receiver are driven by non-coherent statistics over $N$-sample windows: energy/variance estimates for basic NoiseMod and OODN, split-window energies for NC-NoiseMod, and diversity combining across \(M\) slots for TD-NoiseMod. 
To avoid biasing these estimates, the automatic gain control (AGC) should be frozen or slowed within each decision window. 
If this is not possible, it is recommended to record the per-window gain and compensate for it digitally. 
An RF/baseband chain matched to the transmitter’s shaping helps keep the observed process close to white Gaussian noise. 
At startup, and periodically thereafter, it is important to run calibrations to measure the receiver noise floor, correct IQ gain/phase and DC offsets, and refresh decision thresholds from pilot statistics.
In practical terms, a low-cost power-estimation path (squared-magnitude accumulator) feeding a digitally programmable comparator block provides a robust, low-latency realization of the variance test used across these schemes. 
Together, these practices enable synthetic-noise modulations to be deployed without unnecessary complexity or power.

\section{Applications and Integration}\label{Sec:Application}

One of the most promising applications of TherMod lies in battery-less IoT networks. 
By leveraging passive components and energy-harvesting mechanisms, devices can operate without the need for dedicated RF carriers or large batteries. 
This enables periodic data transmission from sensors in hard-to-reach or maintenance-free environments.
Moreover, TherMod offers a viable alternative to backscatter systems, as it does not depend on external ambient RF signals.

TherMod also opens new opportunities for physical-layer security and covert operations. 
The noise-like nature of the transmitted signal makes detection by adversaries difficult, supporting stealth communication in contested or privacy-sensitive environments. 
When combined with RIS, thermal noise can be emitted and modulated without any RF chain, enabling new designs for ultra-low-power and secure transmitters.

Noise-based modulation schemes, such as JEIH-NoiseMod~\cite{Yapici2025}, also provide a pathway for SWIPT. 
By embedding information in the mean of Gaussian noise, these systems can deliver data while maximizing harvested energy. 
The ability to flexibly split time or resources between energy and information makes TherMod systems particularly suited for sustainable and green wireless networks.

Recent advancements demonstrate that noise-based modulation schemes are not limited to ultra-low-power or covert short-range communication, but can be extended to address security and scalability challenges in broader wireless systems.
The hybrid NJM scheme exemplifies this by combining thermal (or artificial) noise modulation with active jamming modulation to counteract intelligent jammers~\cite{Shi} robustly. 
This dual-layer approach provides both low-complexity transmission and resilience to jamming, while allowing dynamic power allocation between the two modes to optimize the performance~\cite{Shi}.

In~\cite{NOMA}, the ND-NOMA architecture explores multi-user access by embedding information into the mean and variance of Gaussian noise, enabling users to share spectrum resources non-orthogonally without requiring high-complexity signal processing, such as successive interference cancellation. 
This is particularly relevant for energy-constrained IoT networks, as it reduces system complexity and supports low-cost deployments. 

\section{Challenges, Future Directions, Cross-Paradigm Synergies and Standardization for Noise-Based Communications} \label{challenges}

\subsection{Key Challenges}

Despite its transformative potential, noise-based communication faces several fundamental challenges that must be addressed before it can be deployed on a wide scale. 
Beyond the core challenges previously identified, recent research highlights additional specific hurdles that require attention.
\begin{itemize}
    \item \textbf{Refined Security Analysis and New Threats:} 
    While the inherent randomness of noise provides a foundation for security, it does not automatically guarantee an unbreakable system. 
    The KLJN secure key exchange system, a cornerstone of noise-based security, has been subject to an ongoing cycle of proposed attacks and defense mechanisms. 
    Attacks have been explored that leverage non-ideal properties in practical implementations, such as tolerance in resistor values, temperature fluctuations, the resistance of interconnecting wires, and deviations from the ideal Gaussian noise distribution. 
    Furthermore, in related quantum communication systems that also exploit randomness for security, noise and the degradation of quantum effects, such as entanglement, pose a major challenge, potentially compromising system functionality. 
    This necessitates that noise-based systems be designed with robust countermeasures against these physical-layer attacks.
    \item \textbf{Practical Implementation and Hardware Imperfections:} 
    Translating theoretical models into practical hardware introduces several non-idealities. 
    For systems like TherMod that switch between resistors, limitations include switching delays, impedance mismatches, and the finite precision of components. 
    The quality of quantum memory is also a crucial factor in quantum entanglement-assisted communication, as stored entangled pairs can lose their properties over time due to noise, leading to errors. 
    For systems using artificial noise generators, ensuring the generated noise maintains the precise statistical properties (e.g., perfect Gaussian distribution) required by theory is challenging, and any imperfection can create security loopholes or performance degradation.
\end{itemize}

\subsection{Future Directions}

Future research in noise-based communication should focus on several promising directions that build upon current foundations while addressing existing limitations.
\begin{itemize}
    \item \textbf{Advanced Protocols for Enhanced Security and Efficiency:} 
    Future work should investigate sophisticated protocols beyond basic schemes.
    For instance, the generalized KLJN protocol utilizes arbitrary resistors rather than just two identical pairs. 
    This flexibility enables real-time compensation of non-ideal component behaviors (e.g., tolerances, wire resistance), thereby improving security in practical deployments. 
    Another concept is Flip-KLJN, a dynamic protocol that flips resistor-bit mappings based on exchanged random bits, allowing all possible noise levels to be used securely and potentially doubling the effective key generation rate compared to the classical KLJN scheme.
    \item \textbf{Integration with Quantum and Advanced Network Concepts:}
    The intersection of noise-based communication and quantum principles represents fertile ground for research. 
    While quantum communication faces noise challenges, advancing techniques show promise. 
    Experiments demonstrate that multi-station quantum networks can be configured to activate non-locality and overcome noise effects. 
    Exploring how such quantum network resilience can inspire classical noise-based systems is a compelling endeavor. 
    Furthermore, the concept of entanglement-assisted communication uses quantum entanglement as a resource to enhance data transmission efficiency and security. 
    Investigating the hybridization of these quantum principles with classical noise-based methods could lead to breakthroughs.
    \item \textbf{Hardware Development and Roadmapping:} 
    A critical direction involves focused development of practical hardware. Current research prototypes, such as those using Johnson noise for passive data transmission, demonstrate low data rates. 
    A key research vector involves optimizing the SNR, increasing the data rate, and extending the range of these systems. 
    Concurrently, developing high-quality quantum memories with longer storage times is essential for making entanglement-assisted communication more viable. 
    Creating a clear roadmap for hardware maturation, from laboratory prototype to industrial standard, is crucial for field growth.
\end{itemize}

\subsection{Cross-Paradigm Synergies and Standardization}

The evolution of noise-based communication will not occur in isolation. 
Seeking synergies with other established and emerging communication paradigms can accelerate the adoption of new approaches.
\begin{itemize}
    \item \textbf{Synergy with Artificial Noise for Physical-Layer Security:} 
    Strong conceptual alignment exists with the broader field of artificial noise for physical-layer security. 
    Artificial noise techniques leverage multiple antennas to generate signals that jam eavesdroppers without affecting legitimate receivers. 
    Research could explore hybrid systems that combine the ultra-low-power benefits of thermal noise modulation with the spatial-beamforming advantages of artificial noise, creating highly secure and efficient communication links for 6G and IoT networks.
    \item \textbf{Standardization and Interoperability:} 
    For any technology to achieve widespread deployment, standardization is paramount. 
    As research matures, efforts must focus on defining standardized modulation formats, frame structures, and medium access control protocols for noise-based systems. 
    This includes establishing how these systems can coexist and interoperate with conventional carrier-based wireless networks (like 802.11 Wi-Fi) that share the same frequency bands.
    Proactive engagement with standards bodies will be crucial for the real-world impact of noise-based communications.
\end{itemize}

\section{Conclusion}
\label{conclusions}

This survey has comprehensively explored noise-based communication, a transformative paradigm that leverages the statistical properties of noise for ultra-low-power, covert, and secure information transfer. 
We have systematically analyzed its core principles, ranging from foundational TherMod to advanced artificial noise schemes (NoiseMod) and their non-coherent and time-diversity variants, as well as their crucial role in secure key exchange via the KLJN protocol.

Our synthesis confirms that noise-based systems offer a uniquely synergistic combination of energy efficiency, hardware simplicity, and inherent physical-layer security, addressing critical demands for future IoT and 6G networks.
Nevertheless, key challenges in robust channel estimation, hardware imperfections, and multi-user scalability remain active research frontiers.

Future progress hinges on developing intelligent, low-complexity detectors, deepening integration with quantum-inspired concepts and RIS, and initiating crucial standardization efforts.
By directly confronting the fundamental trade-offs of next-generation wireless systems, noise-based communication is poised to become a cornerstone technology for enabling ubiquitous, secure, and sustainable connectivity.




\begin{IEEEbiography}[{\includegraphics[width=1in,height=1.25in,clip,keepaspectratio]{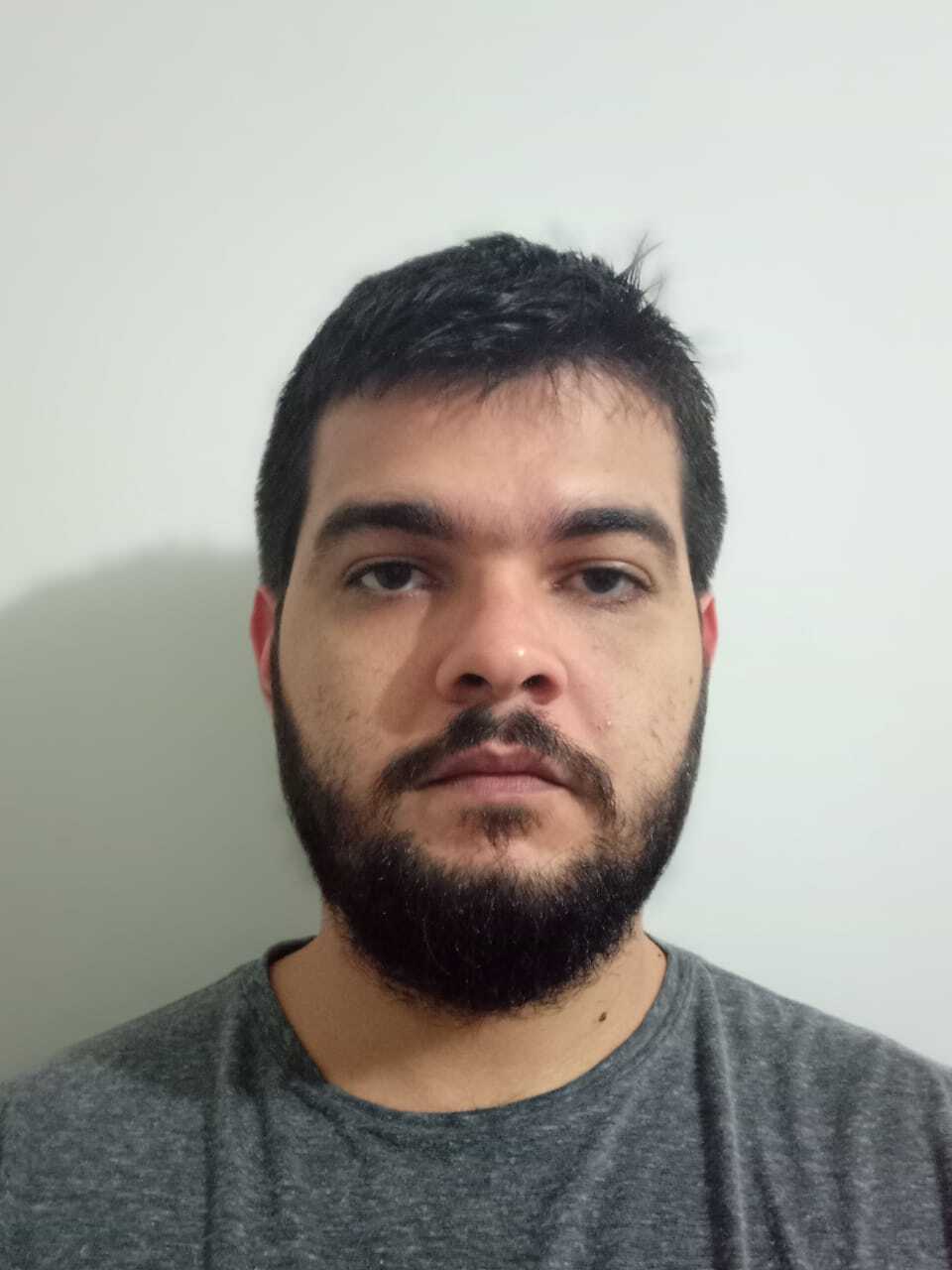}}]{HIGO T. P. SILVA} was born in Nova Floresta, Paraíba, Brazil, in 1993. He received the B.Sc. degree in electrical engineering in 2016 at the Federal University of Paraíba. At the Federal University of Campina Grande, he received his M.Sc. and D.Sc. degrees in electrical engineering in 2018 and 2023, respectively. He is currently an Adjunct Professor at the University of Brasília (UnB). His research interests include signal processing and the channel modeling of wireless communication.
development, and ML.
\end{IEEEbiography}

\begin{IEEEbiography}[{\includegraphics[width=1in,height=1.25in,clip,keepaspectratio]{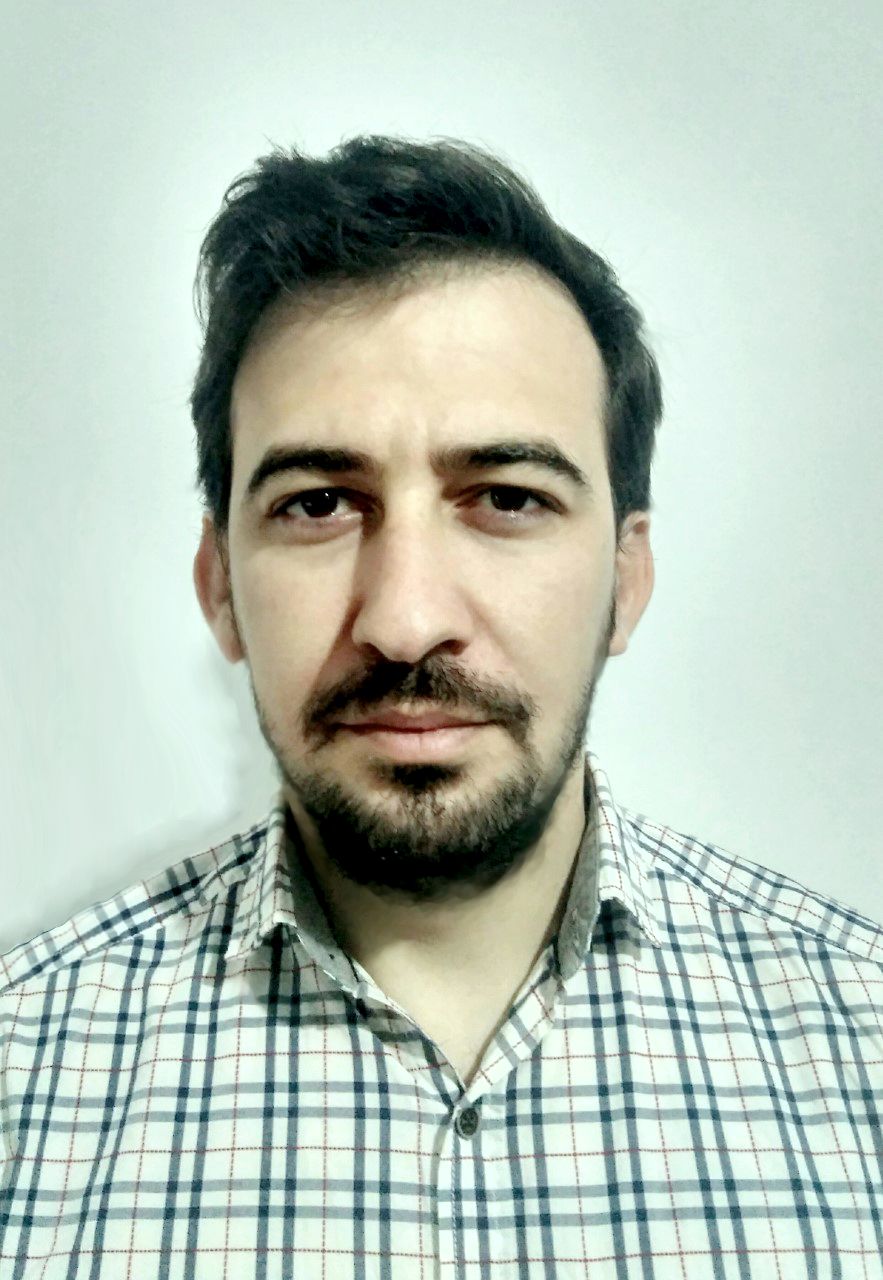}}]{Hugerles S. Silva} (Senior Member, IEEE)  received the B.Sc., M.Sc., and D.Sc. degrees in electrical engineering from the Federal University of Campina Grande, Brazil, in 2014, 2016, and 2019, respectively. Since 2020, he is a Researcher in the Instituto de Telecomunicações (IT) of the University of Aveiro, Portugal. From August to October 2024, he was a Visiting Professor at the Universidad Tecnológica de Panamá, Panama. He is currently an Adjunct Professor at the University of Brasília (UnB). Prof. Silva has been involved in the organizing committee of several conferences and serves as chair of the IEEE Microwave Theory and Technology Society (MTT-S) UnB Student Branch Chapter. He was the recipient of the Latin America Region Young Professional Award from the IEEE Communications Society in 2022. He was the general coordinator of the 2023 Workshop on Communication Networks and Power Systems (WCNPS 2023). His main research interests include digital signal processing and wireless channel modeling. He is a senior member of IEEE and a reviewer for several journals. He is the author of more than 80 publications in scientific journals and international conferences.
\end{IEEEbiography}

\begin{IEEEbiography}[{\includegraphics[width=1in,height=1.25in,clip,keepaspectratio]{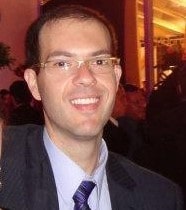}}]{FELIPE A. P. de FIGUEIREDO } received the B.S. and M.S. degrees in
telecommunications from the Instituto Nacional de Telecomunicações
(INATEL), Minas Gerais, Brazil, in 2004 and 2011, respectively. He
received his Ph.D. from the State University of Campinas (UNICAMP),
Brazil, in 2019. He has worked in the Research and Development of
telecommunications systems for over fifteen years. His research
interests include digital signal processing, digital communications,
mobile communications, MIMO, multicarrier modulations, FPGA
development, and ML.\end{IEEEbiography}

\begin{IEEEbiography}[{\includegraphics[width=1in,height=1.25in,clip,keepaspectratio]{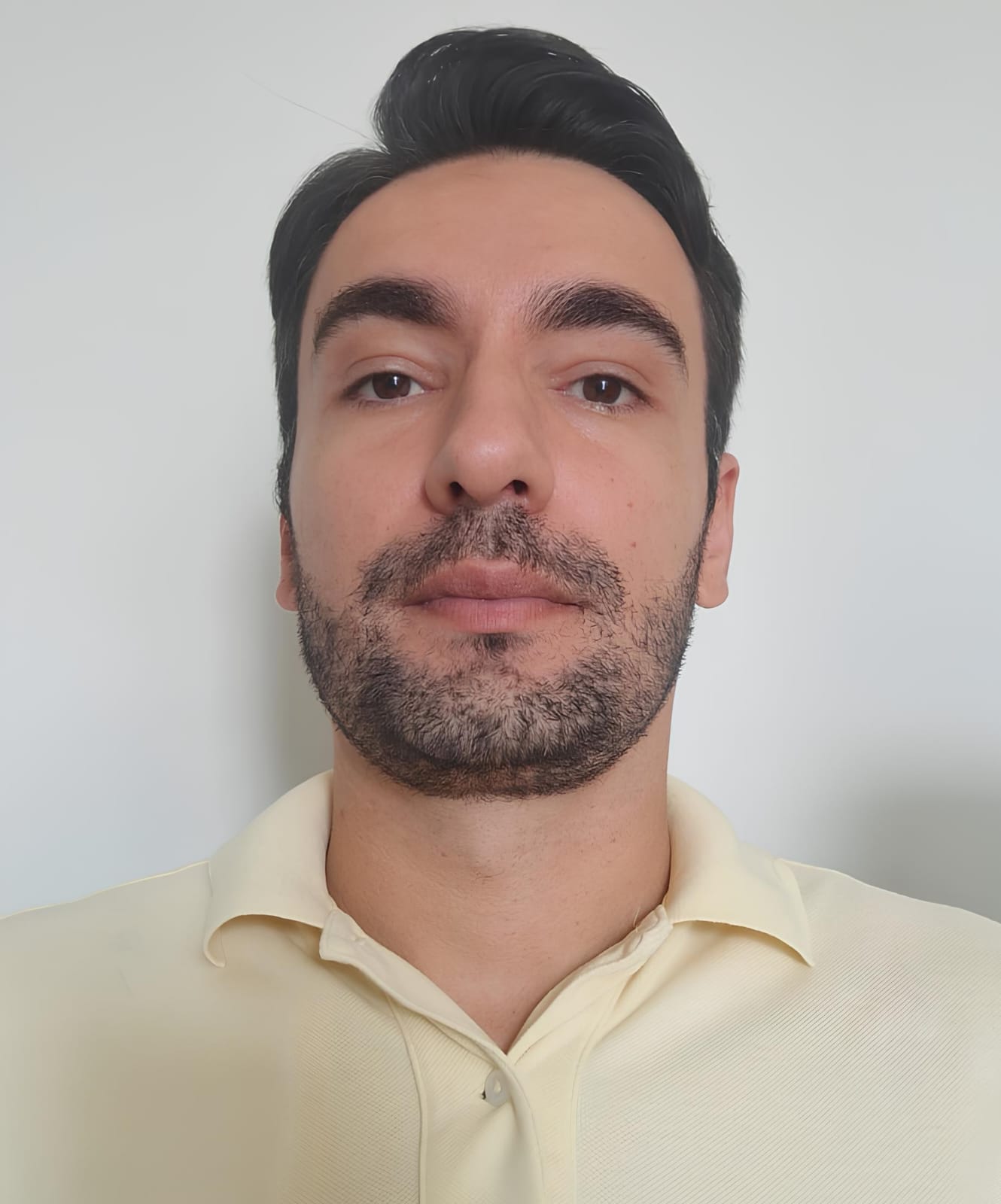}}]{ANDRÉ A.\ DOS ANJOS} received his Ph.D. from the University of Campinas in 2021 and worked for over 11 years at the Inatel Competence Center, developing telecommunications and digital signal processing projects. He is now a professor of Electronic and Telecommunications Engineering at the Federal University of Uberlândia, Campus Patos de Minas, Brazil, with research interests in wireless communications, fading channel modeling and simulation, nonlinear systems, and spectrum sensing.\end{IEEEbiography}

\begin{IEEEbiography}[{\includegraphics[width=1in,height=1.25in,clip,keepaspectratio]{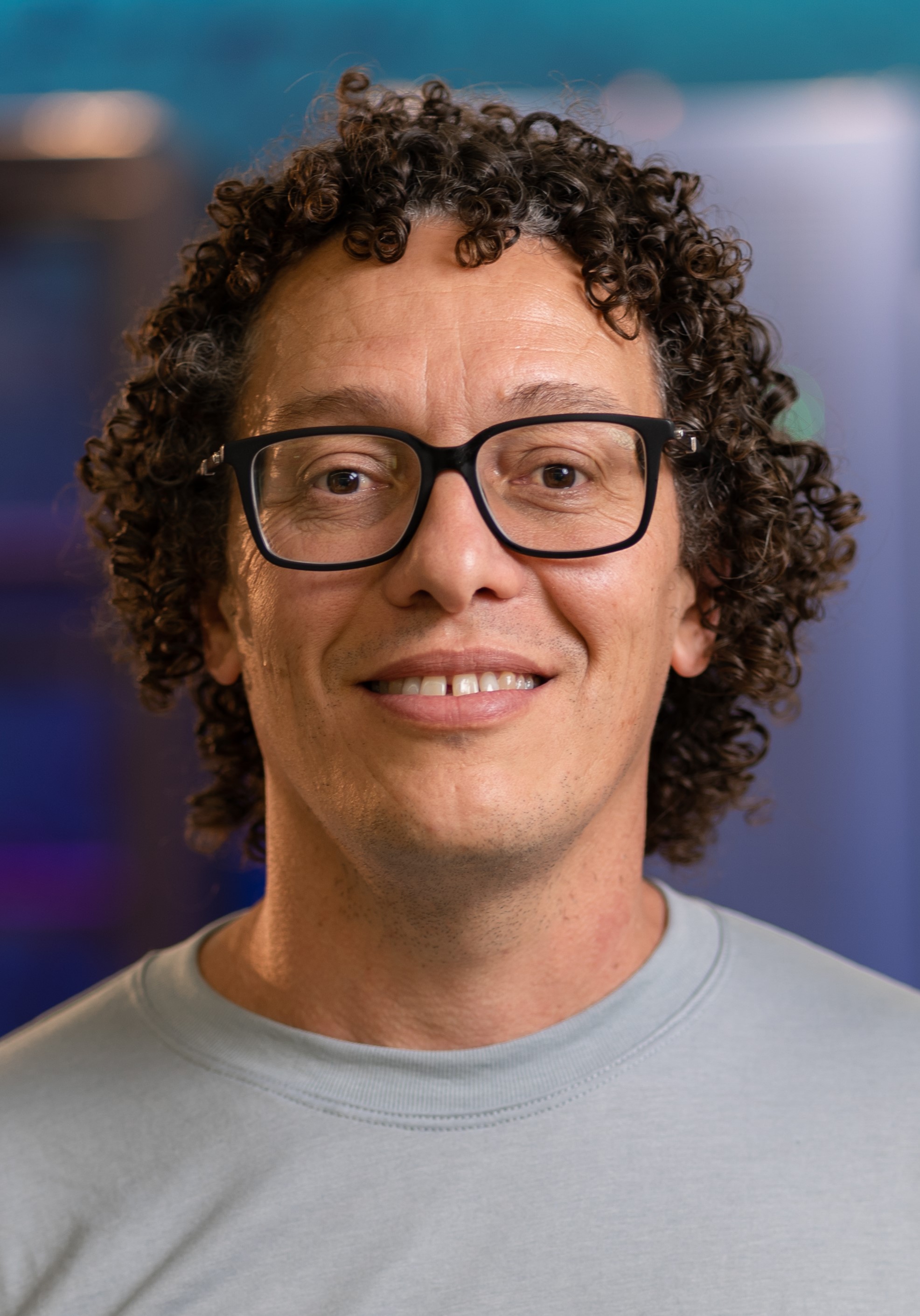}}]{Rausley Adriano Amaral de Souza} (Senior Member, IEEE) received the B.S.E.E.
and M.Sc. degrees from the National Institute of Telecommunication (INATEL), Brazil,
in 1994 and 2002, respectively, and the Ph.D.
degree from the State University of Campinas
(UNICAMP), Campinas, Brazil. Before joining
the Academy, he worked in the industry. He joined
INATEL in 2002, where he is a Full Professor. His research interests include wireless
communications.
\end{IEEEbiography}

\EOD

\end{document}